\documentclass[letterpaper, 10 pt, conference]{ieeeconf}

\IEEEoverridecommandlockouts     
\usepackage{algorithm}
\usepackage{array}
\usepackage{stfloats}
\usepackage{tabularx}
\usepackage{url}
\usepackage{verbatim}
\usepackage{subcaption}
\usepackage{cite}
\usepackage{stackengine}
\usepackage{booktabs}
\usepackage{amsmath,amssymb,amsfonts}
\usepackage{algorithmic}
\usepackage{graphics} 
\usepackage{epsfig} 
\usepackage[absolute,overlay]{textpos}
\usepackage{textcomp}
\usepackage{xcolor}
\usepackage{amsthm}
\usepackage{bm}
\usepackage{comment}
\usepackage[noabbrev]{cleveref}
\usepackage[font=small,labelfont=bf]{caption}

\newtheorem{definition}{Definition}

\newtheorem{theorem}{Theorem}

\renewcommand{\ddddot}[1]{%
  {\mathop{\kern\z@#1}\limits^{\vbox to-1.4\ex@{\kern-\tw@\ex@
   \hbox{\normalfont....}\vss}}}}

\DeclareMathOperator*{\argmin}{arg\,min}

\DeclareMathOperator{\sgn}{sgn}

\def\BibTeX{{\rm B\kern-.05em{\sc i\kern-.025em b}\kern-.08em
    T\kern-.1667em\lower.7ex\hbox{E}\kern-.125emX}}

\begin{document}
\title{Control Barrier Functions for Shared Control  and Vehicle Safety}
\author{James Dallas$^{1*}$, John Talbot$^{1}$, Makoto Suminaka$^{1}$, Michael Thompson$^{1}$, 
Thomas Lew$^{1}$, \\ G\'abor Orosz$^{2}$, and John Subosits$^{1}$
\thanks{$^{1}$%
Toyota Research Institute, Los Altos, CA, USA. Contact emails:
\{james.dallas, john.talbot, makoto.suminaka, michael.thompson, thomas.lew, john.subosits\}@tri.global.  
}%
\thanks{$^{2}$%
	University of Michigan, Ann Arbor, MI, USA. Contact email: orosz@umich.edu.
}
\thanks{*%
	Corresponding author James Dallas. james.dallas@tri.global.
}%
}
\maketitle

\begin{textblock*}{16cm}(3cm,1cm) 
\centering
   \noindent \textit{This is the author's accepted manuscript version of the paper:  J. Dallas, J. Talbot, M. Suminaka, M. Thompson, T. Lew, G. Orosz, and J. Subosits, "Control Barrier Functions for Shared Control and Vehicle Safety", American Control Conference, 2025.}
\end{textblock*}

\begin{abstract}
This manuscript presents a control barrier function based approach to shared control for preventing a vehicle from entering the part of the state space where it is  unrecoverable. 
The maximal phase recoverable ellipse is presented as a safe set in the sideslip angle--yaw rate  phase plane where the vehicle's state can be maintained.
An exponential control barrier function is then defined on the maximal phase recoverable ellipse to promote safety.
Simulations demonstrate that this approach enables safe drifting,
that is, driving at the handling limit without spinning out. 
Results are then validated for shared control drifting with an experimental vehicle in a closed course. 
The results show the ability of this shared control formulation to maintain the vehicle's state within a safe domain in a computationally efficient manner, even in extreme drifting maneuvers.
\end{abstract}
    


\section{Introduction}

Expanding the operational range of (semi-)autonomous vehicles can increase the feasible action space available to safety systems in emergencies. 
Indeed, current safety systems, such as electronic stability control (ESC), operate on a small subset in the state space where open-loop stability is preserved. 
Yet, skilled drivers can precisely maneuver a vehicle in a continued state of oversteer when drifting. Having advanced safety systems capable of similar levels of maneuverability could improve safety in practical scenarios such as a vehicle oversteering when encountering a snow patch. 

One successful approach for autonomous vehicles executing maneuvers at and beyond the stable handling limits in racing and drifting is nonlinear model predictive control (NMPC) \cite{dallas2023hierarchical, LaurenseThesis, Goh2023}. 
Recent works have demonstrated how these control techniques can amplify the driver's abilities in a shared control setting through parallel autonomy and keep the system safe even when the vehicle is operating in extreme situations \cite{talbot2023shared, SMCDrift}. 
While these works have shown the ability to control a vehicle at the handling limits, 
transitioning these methods from research to broader deployment remains challenging due to potentially limited onboard compute on broadly available production vehicles, and  
due to local minima and infeasibility of nonconvex optimization. 

Control barrier functions (CBFs) show particular promise for ensuring safety in a computationally efficient manner as they can be formulated through quadratic programming (QP) \cite{ames2019control}. 
By imposing forward invariance of a set in state space, safety can be guaranteed \cite{XU201554, Ames2017, ECBF, cbfcomp, CBFMPC}.
Unifying CBFs with control Lyapunov funcions (CLFs) can enable safety and performance objectives to be met simultaneously. 
Notably, a QP can ensure that safety is prioritized through the CBF while stability is achieved through the CLF \cite{ames2019control}. 
In the automotive domain, such approaches have been demonstrated for guaranteeing safety via minimal intervention for adaptive cruise control (ACC), lane keeping \cite{XU201554, Ames2017, AVEC2024}, and emergency braking of connected automated vehicles (CAVs) \cite{CAVCBF}.

While these works have demonstrated success in ensuring safety in simple driving scenarios, they do not address the problem of ensuring safety at and beyond the stable handling limits that would require one to account for the strong nonlinearities in vehicle dynamics.
Also, these works do not consider the problem of shared control. 
Racing and drifting are particularly interesting examples to probe CBF approaches for vehicle safety as the vehicle is operating right at the edge of its capabilities and is at the verge of losing control authority.

\begin{figure}[t]
    \centering
    \includegraphics[width=1\columnwidth, trim={0 9cm 0 9cm},clip]{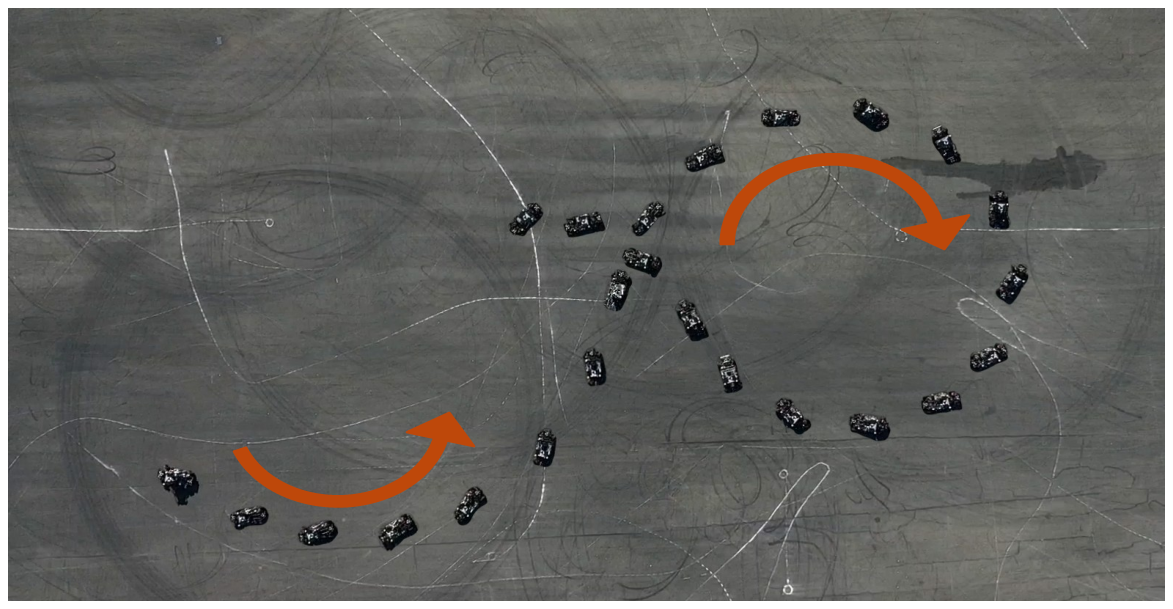}
    \caption{Segment of figure 8 maneuver from experimental validation with the CBF based controller turned on.}
    \label{fig:exp_overlay}
    \vspace{-1.5em}
\end{figure}

This work proposes an approach to mitigate a vehicle violating safety bounds that would lead to unrecoverable motion. 
The maximal phase recoverable ellipse (MPREl) is presented as a safe set in the sideslip angle--yaw rate phase plane where the vehicle remains in a recoverable state. Specifically, the MPREl extends upon \cite{goh2019automated} by applying a smooth approximation to the maximal phase recoverable envelope and developing a closed form, analytical solution.
Next, in contrast to previous approaches developed for racing and drifting, a control barrier function is placed on this ellipse to filter driver commands in a shared control setting. 
In order to handle the higher relative degree of the control system, an exponential control barrier function (ECBF) is utilized in the safety filter.
We demonstrate the utility of this approach with full scale shared control drifting experiments where the vehicle safely operates in the open loop unstable region; see Fig.~\ref{fig:exp_overlay}. 

The rest of the paper is structured as follows. 
Sec.~\ref{sec:background} provides mathematical background on CBFs and states our main safety result. 
The vehicle dynamics, the maximum phase recoverable ellipse based CBF design, and the quadratic program are introduced in Sec.~\ref{sec:approach}. 
The experimental validation with supporting numerical simulations are presented in Sec.~\ref{sec:exp}. 
Finally concluding remarks are in Sec.~\ref{sec:conc}.

\section{Background on Control Barrier Functions}\label{sec:background}

Barrier functions have drawn interest to enable the design of controllers that certify safety of a system through conditions on forward invariance of a given set. 
Here we give a brief overview while more details can be found in \cite{ames2019control, cohen2024safety}. 

Consider a system with control affine dynamics
\begin{flalign}\label{eq:affine}
    &\dot{x} = f(x) + g(x)u,
\end{flalign}
with state ${x \in \mathbb{R}^n}$, input ${u \in \mathbb{R}^m}$, and locally Lipschitz functions $f(x)$ and $g(x)$.
The notion of safety can be characterized through forward invariance of a set in state space for a dynamical system.
\begin{definition}\label{def:invar}
    Forward invariance \cite{ames2019control}: The set ${\mathcal{S} \subset \mathbb{R}^n}$ is forward invariant if  ${x(0) \in \mathcal{S} \Rightarrow x(t) \in \mathcal{S},  \forall t \geq 0}$ for the solutions of \eqref{eq:affine}.
    Then we say that the set $\mathcal{S}$ is safe w.r.t.~\eqref{eq:affine}. 
\end{definition}
Specifically, we consider the set $\mathcal{S}$ to be the 0-superlevel set of a continuously differentiable function ${h: \mathbb{R}^n \rightarrow \mathbb{R}}$:
\begin{flalign}\label{eq:barrier}
    \mathcal{S} &=  \{x \in \mathbb{R}^n : h(x) \geq 0\}. 
\end{flalign}
Therefore the forward invariance of the set $\mathcal{S}$ can be characterized by maintaining the non-negativity of the function $h$.
That is, to certify safety, one needs to show that ${h(x(0)) \geq 0  \Rightarrow  h(x(t)) \geq 0, \forall t \geq 0}$.
This leads to the following definition.
\begin{definition}\label{def:CBF} 
Control barrier function (CBF) \cite{ames2019control}: The continuously differentiable function 
${h: \mathbb{R}^n \rightarrow \mathbb{R}}$ is a
CBF for \eqref{eq:affine} on $\mathcal{S}$ defined by \eqref{eq:barrier}, if there exists ${\alpha_0>0}$ such that ${\forall x \in \mathcal{S}}$
\begin{equation}\label{eq:CBF_condition}
\sup_{u\in \mathbb{R}^m} \big[\dot{h}(x,u) \big] =
\sup_{u\in \mathbb{R}^m} \big[L_f h(x) + L_g  h(x) u\big] > - \alpha_0 \, h(x),
\end{equation}
where ${L_f h(x) = \nabla h(x)  f(x)}$ and ${L_g h(x) = \nabla h(x)  g(x)}$ are the Lie derivatives of $h$ along $f$ and $g$.
\end{definition}
To be more general, one may use a class-$\mathcal{K}$ function of $h$ on the right hand side instead of the linear function with gradient $\alpha_0$, but in most practical applications the above setup is adequate.
With Definition \ref{def:CBF} we can state a theorem which can be used to synthesize safe controllers.
\begin{theorem}\label{thm:safety}
\cite{ames2019control}:
If $h$ is a CBF for \eqref{eq:affine} on $\mathcal{S}$ defined by~\eqref{eq:barrier}, then any locally Lipschitz continuous controller ${\xi : \mathbb{R}^n \to \mathbb{R}^m}$, with ${u=\xi(x)}$ satisfying
\begin{equation}\label{eq:safety_cond_control}
\dot{h}(x,u)= L_f h(x) + L_g  h(x) u \geq - \alpha_0 \,h(x),
\end{equation}
${\forall x \in \mathcal{S}}$ renders set $\mathcal{S}$ safe w.r.t.~\eqref{eq:affine}.
\end{theorem}

Condition \eqref{eq:safety_cond_control} can be used as a constraint when synthesizing controllers via quadratic programming (QP) if  ${L_g  h(x) \neq 0}$. 
However, in many practical cases this condition does not hold.
Such systems still admit controllers leveraging CBFs to ensure forward invariance, and are based on the concept of relative degree.
\begin{definition}\label{def:reldeg}
The $k$ times continuously differentiable function ${h: \mathbb{R}^n \rightarrow \mathbb{R}}$ has relative degree ${k\geq2}$ for \eqref{eq:affine} if ${\forall x \in \mathbb{R}^n}$ we have ${L_gh(x)=L_gL_fh(x) = \ldots = L_gL_f^{k-2}h(x) = 0}$ and ${L_gL_f^{k-1}h(x) \neq 0}$.
\end{definition}
For systems with dynamics \eqref{eq:affine}  with a higher relative degree, safety  conditions of type \eqref{eq:safety_cond_control} can be imposed using an exponential CBF (ECBF) \cite{ames2019control, ECBF, XIAO2021109592}. 
Define the system
\begin{flalign}
   &\Dot{\eta}(x) = F\eta(x) + G\mu, \nonumber 
   \\
   &h(x) = C\eta(x),
\end{flalign}
where
\small
\begin{flalign}
F &=
\begin{bmatrix}
0 & 1 & 0 & \ldots & 0\\
0 & 0 & 1 & \ldots & 0\\
\vdots & \vdots & \vdots & \ddots \\
0 & 0 & 0 & \ldots & 1\\
0 & 0 & 0 & \ldots & 0
\end{bmatrix} , 
G =
\begin{bmatrix}
0 \\
\vdots\\
0 \\
1
\end{bmatrix}, 
\eta(x) =
\begin{bmatrix}
h(x) \\
L_f(h(x))\\
\vdots\\
L_f^{k-1}(h(x))
\nonumber
\end{bmatrix},
\\
C &=
\begin{bmatrix}
1 & 0 & \cdots & 0 
\end{bmatrix}. 
\end{flalign}
\normalsize
Furthermore, define the sets ${\mathcal{S}_i\subset\mathbb{R}^n}$ as 0-superlevel sets of the functions 
${\nu_i(x) : \mathbb{R}^n \rightarrow \mathbb{R}}$ as
\small
\begin{flalign} \label{eq:hoset}
&\nu_0(x) = h(x), ~~~~~~~~~~~~~~~~~~~~~~~~~~ \mathcal{S}_0=\{x \in \mathbb{R}^n: \nu_0(x) \geq 0\}, \nonumber 
\\
&\nu_1(x) = \Dot{\nu}_0(x) + \alpha_0 \nu_0(x), ~~~~~~~~~~~ \mathcal{S}_1=\{x \in \mathbb{R}^n: \nu_1(x) \geq 0\}, \nonumber 
\\
& ~~~~~~~~~~~~~~~~~~~~~~~~~~~~~~~~~\vdots 
\\
&\nu_k(x) = \Dot{\nu}_{k-1}(x) + \alpha_{k-1} \nu_{k-1}(x), ~ \mathcal{S}_k=\{x \in \mathbb{R}^n: \nu_k(x) \geq 0\}, \nonumber
\end{flalign}
\normalsize
where ${\mathcal{S}_0=\mathcal{S}}$ by definition. 
The constants ${-\alpha_i\in\mathbb{R}}$, ${i=0,\ldots,k-1}$ are the (real) roots of the characteristic polynomial of ${F-GP^\top}$. 
That is, we have 
\begin{flalign}\label{eq:char}
&(\lambda+\alpha_0)(\lambda+\alpha_1)\cdots(\lambda+\alpha_{k-1}) 
\\
&= \lambda^k + p_{k-1} \lambda^{k-1} + \cdots + p_1 \lambda + p_0 = 0,\nonumber
\end{flalign}
with coefficients ${p_i\in\mathbb{R}}$, ${i=0,\ldots,k-1}$. 
These coefficients can be used to construct the state feedback 
$$
{\mu = -P^\top\eta(x)}, \quad\text{where }P^\top = [\, p_0 \ p_1 \ \cdots \ p_{k-1}\,].
$$ 

Then, the ECBF is defined as follows.
\begin{definition}\label{def:HRCBF} 
Exponential control barrier function (ECBF) \cite{ames2019control}:
Given a set ${\mathcal{S} \subset \mathbb{R}^n}$ defined as the 0-superlevel set of a $k$ times continuously differentiable function ${h : \mathbb{R}^n \rightarrow \mathbb{R}}$, then $h$ is an ECBF if there exists ${P \in \mathbb{R}^k}$ such that ${\forall x \in {\rm Int}(\mathcal{S})}$, 
\begin{equation}
     \sup_{u \in \mathbb{R}^m} \big[ L^k_fh(x) + L_gL_f^{k-1}h(x)u \big] > -P^\top\eta(x), 
\end{equation}
results in ${h(x(t)) \geq C\mathrm{e}^{(F-GP^\top)t}\eta(x(0)) \geq 0}$ whenever ${h(x(0)) \geq 0}$.
\end{definition}

The gain matrix $P$ can be chosen to satisfy the ECBF conditions in Definition \ref{def:HRCBF} using the following result.
\begin{theorem}\label{def:ERCBF}\cite{ames2019control, ECBF}:
Suppose ${P \in \mathbb{R}^k}$ is chosen such that the control system ${F-GP^\top}$ has negative real eigenvalues which satisfy 
${-\alpha_i \leq \Dot{\nu}_i(x(0)) / \nu_i(x(0))}$ for ${i=0,\ldots,k-1}$. Then, ${\mu \geq -P^\top\eta(x)}$ guarantees that $h(x)$ is an ECBF.
\end{theorem}
We then state the following theorem which leverages arguments in \cite{molnar2023safety} to show the forward invariance of the set ${\mathcal{S} \cap \mathcal{S}_1 \cap \cdots \cap \mathcal{S}_{k-1}}$. 
\begin{theorem}\label{thm:esafety}
\textit{(Main Result)} If $h$ is a ECBF for \eqref{eq:affine} with sets $\mathcal{S}_i$, ${i=0,\ldots,k}$ defined by~\eqref{eq:hoset}, then any locally Lipschitz continuous controller ${\xi : \mathbb{R}^n \to \mathbb{R}^m}$, with ${u=\xi(x)}$ satisfying
\begin{equation}\label{eq:esafety_cond_control}
L^k_fh(x) + L_gL_f^{k-1}h(x)u \geq -P\eta(x), 
\end{equation}
${\forall x \in \mathcal{S}_{k-1}}$ renders set ${\mathcal{S} \cap \mathcal{S}_1 \cap \cdots \cap \mathcal{S}_{k-1}}$ safe w.r.t.~\eqref{eq:affine}.
\end{theorem}
\begin{proof}
If a controller satisfies the condition \eqref{eq:esafety_cond_control}
then it renders ${S}_{k-1}$ forward invariant according to Theorem~\ref{def:ERCBF}. 
Then, following the argument in Corollaries 1 and 2 in \cite{molnar2023safety}, ${{S}_{k-2} \cap {S}_{k-1}}$ is forward invariant. 
Using the same logic yields that ${{S}_{k-3} \cap {S}_{k-2} \cap {S}_{k-1}}$ is forward invariant and
applying this approach successively yields the forward invariance of
${\mathcal{S} \cap \mathcal{S}_1 \cap \cdots \cap \mathcal{S}_{k-1}}$. 
\end{proof}

Condition \eqref{eq:esafety_cond_control} can be utilized as a constraint in a QP when synthesizing controllers as will be applied below for the vehicle model.
Again, one may use class-$\mathcal{K}$ functions instead of the linear ones with gradients ${\alpha_i\in\mathbb{R}}$, ${i=0,\ldots,k-1}$ and construct a so-called extended CBF \cite{Xiao2019,cohen2024safety}. 
This may provide one with more flexibility in control design but the linear functions already give formal safety guarantees. 
Below we demonstrate that using an ECBF can indeed render the vehicle safe in real experimental scenarios.

\section{Control Barrier Functions for Vehicle Safety}\label{sec:approach}

Here we present the nonlinear vehicle model used for control design. 
We introduce the maximum phase recoverable ellipse on the sideslip angle--yaw rate phase plane which the control barrier function is designed with.
Finally, a quadratic programming safety filter is designed.

\begin{figure}[t]
    \includegraphics[width=\columnwidth]{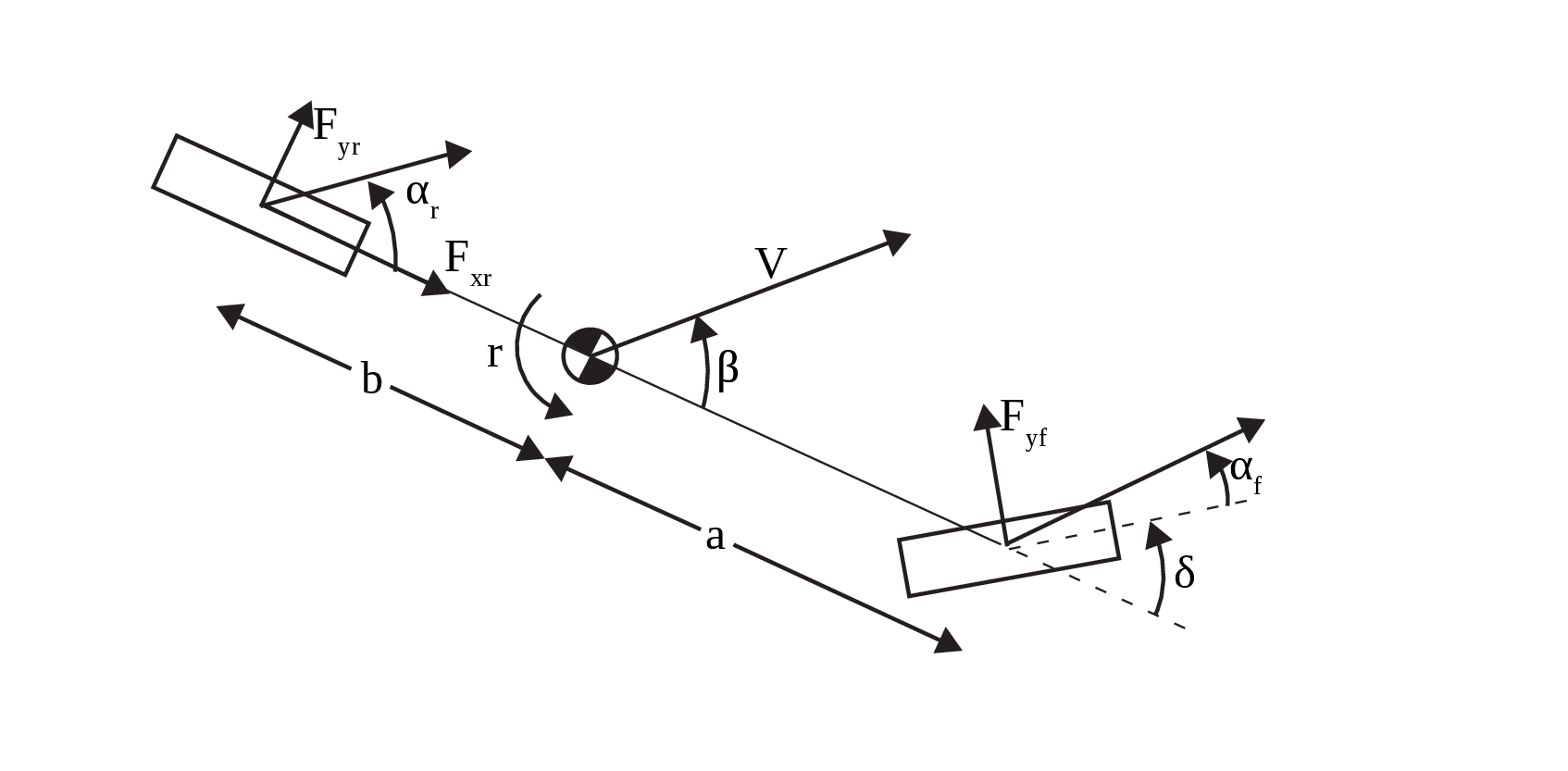}
    \centering
    \caption{Bicycle model used in this paper.}
    \label{fig:vehicle model}
\end{figure}

\subsection{Vehicle Dynamics}

To construct a candidate CBF that can operate in extreme safety maneuvers, a suitable vehicle model is needed. 
Works \cite{LaurenseThesis, Goh2023} have shown that a 3 degree-of-freedom single track model (depicted in Fig.~\ref{fig:vehicle model}) describes vehicle behavior adequately even in extreme situations such as racing and drifting. 
Since we focus on vehicle handling, we omit the configuration coordinates and describe the vehicle using the velocity states, namely, the yaw rate $r$, the slideslip angle $\beta$ of the center of mass, and the speed $V$ of the center of mass.
The steering angle $\delta$ and the torque $\tau$ applied at the rear wheel are also added to the system state by assuming that we command the steering rate $\dot\delta$ and the torque rate $\dot\tau$.

Defining the state ${x=[\,r\ \beta\ V\ \delta\ \tau\,]^\top}$ and the input ${u=[\, \dot\delta\ \dot\tau\,]^\top}$
yields the control affine form \eqref{eq:affine} with functions
\begin{flalign}\label{eq:r}
    f(x) &= 
    \begin{bmatrix}
    \frac{a ( F_{x, {\rm f}}\sin\delta + F_{y, {\rm f}}\cos\delta ) - b F_{y, {\rm r}}}{I_z}      
    \\
    \frac{F_{x, {\rm f}}\sin(\delta - \beta) + F_{y, {\rm f}}\cos(\delta - \beta) -F_{x, {\rm r}}\sin\beta + F_{y, {\rm r}}\cos\beta}{mV}-r 
    \\
    \frac{F_{x, {\rm f}}\cos(\delta - \beta) - F_{y, {\rm f}}\sin(\delta - \beta) + F_{x, {\rm r}}\cos\beta + F_{y, {\rm r}}\sin\beta}{m} 
    \\
    0 
    \\
    0 
    \end{bmatrix}, \nonumber 
\\
    g(x) &= 
    \begin{bmatrix}
    0 & 0 & 0 & 1 & 0
    \\
    0 & 0 & 0 & 0 & 1 
    \\
    \end{bmatrix}^{\top},
\end{flalign}
where $a$ and $b$ are the distances between the center of mass and front and rear axles, $m$ is the mass of the vehicle, and $I_z$  is the moment of inertia about center of mass.

The lateral tire forces are given by the tire model in \cite{LaurenseThesis}:
\begin{flalign}\label{eq:lateral}
    F_{y} =  
    \begin{cases}
    - C_{\rm c} \tan\alpha 
    + \frac{C_{\rm c}^2}{3F_{y, {\rm max}}} |\tan\alpha|\tan\alpha  
    & \hspace{-3mm} -  \frac{C_{\rm c}^3}{27 F_{y, {\rm max}}^2} \tan^3\alpha 
    \\
    &\quad {\rm if} \quad  |\alpha| < \alpha_\mathrm{sl},
    \\
    -F_{y, {\rm max}}\sgn\alpha & \quad {\rm if} \quad |\alpha| > \alpha_\mathrm{sl},
\end{cases}
\end{flalign}

\noindent
with ${\alpha_\mathrm{sl} = {\rm atan}\left(\frac{3F_{y, {\rm max}}}{C_{\rm c}}\right)}$ and ${F_{y, {\rm max}} = \sqrt{(\mu F_z)^2-\gamma F_x^2}}$. 
Here $C_{\rm c}$ is the cornering stiffness, $\mu$ is the coefficient of friction, $F_z$ is the normal force, while ${\gamma = 0.99}$ is a tuning parameter to ensure numeric stability as longitudinal force $F_x$ approaches the friction limit.
Note that \eqref{eq:lateral} is identical to the brush model in \cite{oh2021handling} when considering that the sticking and sliding friction coefficients are equal (${\mu_0=\mu}$).

The normal forces for the front and the rear tires are calculated based on the static weight distribution:
\begin{equation}\label{eq:vertical}
F_{z,{\rm f}} = \frac{mgb}{a+b}, \quad F_{z,{\rm r}} = \frac{mga}{a+b}.
\end{equation}
Considering rear wheel drive, the longitudinal forces are
\begin{equation}\label{eq:longitudinal}
F_{x,{\rm f}} = 0, \quad F_{x,{\rm r}} = \frac{\tau}{r_{\rm w}},
\end{equation}
where $\tau$ is wheel torque and $r_{\rm w}$ is the wheel radius.
Finally, the front and rear slip angles can be calculated from the vehicle kinematics, namely, from the velocity of the wheel centers as
\begin{equation}\label{eq:slips}
\begin{split}
    \alpha_\mathrm{f} &= \arctan\left(\frac{V \sin\beta+ar}{V \cos\beta}\right) -\delta ,
    \\
    \alpha_\mathrm{r} &= \arctan\left(\frac{V\sin\beta-br}{V\cos\beta}\right) .
\end{split}    
\end{equation}

Observe that \eqref{eq:affine},\eqref{eq:r},\eqref{eq:lateral},\eqref{eq:vertical},\eqref{eq:longitudinal},\eqref{eq:slips} give a highly nonlinear set of equations which may possess multiple equilibria making the control design a very challenging task.

\subsection{Maximum Phase Recoverable Ellipse and CBF Design}

\begin{figure*}[t]
\vspace{0.5em}
    \centering
     \begin{subfigure}[t]{0.4\textwidth}
         \includegraphics[scale = 0.35,trim={6cm 1cm 3cm 1cm}]{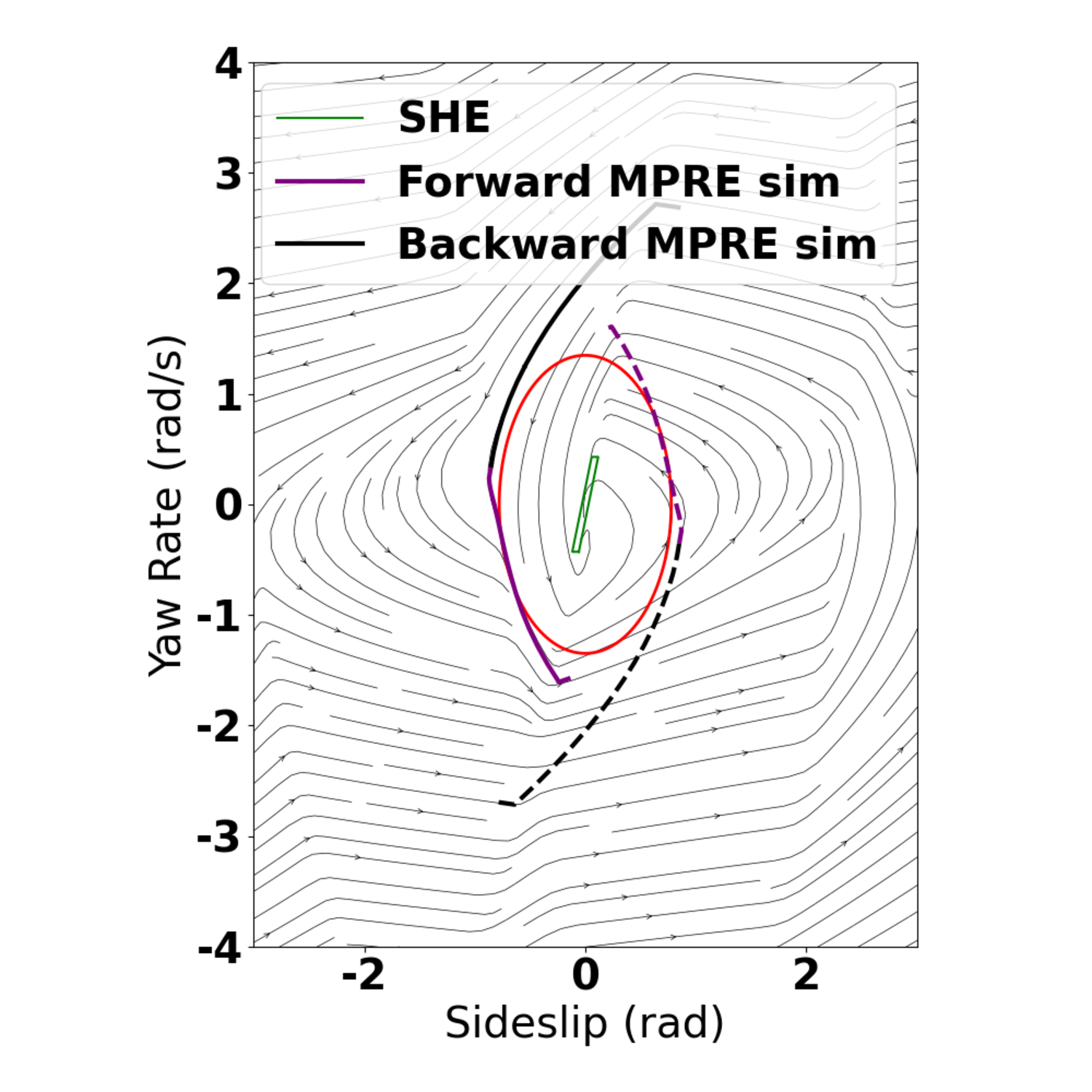}
         \centering
         \caption{Negative steering}
         \label{fig: pp_neg}
     \end{subfigure}
     \begin{subfigure}[t]{0.4\textwidth}
         \centering
         \includegraphics[scale = 0.35, trim={3cm 1cm 2cm 1cm}]{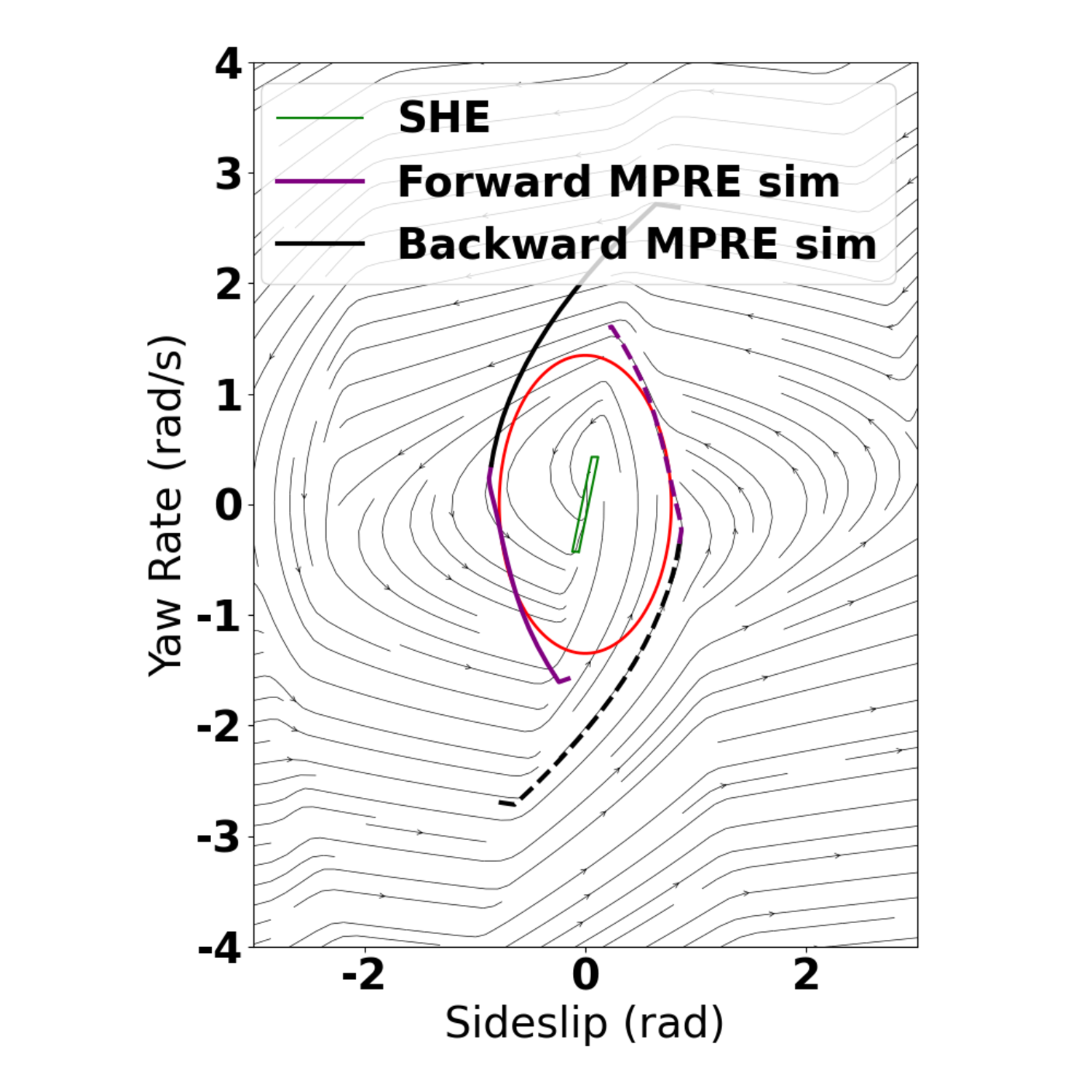}
         \caption{Positive steering.}
         \label{fig: pp_pos}
     \end{subfigure}

    \caption{Phase portrait for constant speed ${V=7\, \mathrm{m/s}}$ when applying the steering angle ${\delta= \delta_{\rm max}= 0.71\, \mathrm{rad}}$ and zero axle torque ${\tau=0}$ for the MPRE (black and purple) and the MPREl (red).}
    \label{fig:phase_portrait}
    \vspace{-1.5em}
    
\end{figure*}

The CBF candidate is defined on the maximum phase recovery ellipse (MPREl) which is constructed as a subset of the maximum phase recovery envelope (MPRE) defined in \cite{goh2019automated}. 
While other representations could be considered, such as the intersection of four half planes \cite{CombinedCBF}, the MPREl captures a large portion of the MPRE in a single constraint.
Briefly, the MPRE bounds the set where a vehicle remains in an open loop unstable yet still recoverable state. 
Beyond the MPRE, the vehicle loses control authority and can no longer be stabilized, leading to a spin out. 
In constructing the set, it is useful to visualize the phase portrait corresponding to maximum countersteer ${\delta = \pm \delta_{\rm max}}$ and zero axle torque ${\tau=0}$, as this is an input that captures nearly all recoverable states \cite{goh2019automated} by maximizing the tire force needed to balance the yaw moment.  
Fig.~\ref{fig:phase_portrait} depicts the phase portrait for negative countersteer (panel (a)) and positive countersteer (panel (b)) at zero axle torque.
An elliptical approximation, namely the MPREl, is also shown (red). 

As in \cite{DallasMPRP}, the MPRE is constructed for desired sideslip angles ${\beta = \pm \beta_{\rm max}}$ by obtaining a point on the beta nullcline ${\dot\beta = 0}$:
\begin{equation} \label{MPRE_point}
\begin{cases}
    r = \frac{\mu  F_{z, {\rm f}}\cos(-\delta_{\rm max}-\beta_{\rm max}) + \mu  F_{z, {\rm r}} \cos\beta_{\rm max}}{mV} & \text{if } \delta \le 0, 
    \\
    r = \frac{-\mu  F_{z, {\rm f}}\cos(\delta_{\rm max}-\beta_{\rm max}) - \mu F_{z, {\rm r}} \cos\beta_{\rm max}}{mV} & \text{if } \delta \ge 0,
    \end{cases}
\end{equation} 
where we considered that at the handling limit ${F_{y} = \mu F_z}$.
Then the MPRE is obtained by forward (purple) and reverse (black) simulating the dynamics.
Then fitting the largest ellipse within these trajectories gives the MPREl (red); see Fig.~\ref{fig:phase_portrait}. Notably, trajectories are observed to converge even outside of the ellipse thus making it a conservative estimate should the barrier be breached, e.g., due to model mismatch or actuation limits.

To mitigate a loss of control, the states are constrained to remain inside the MPREl. 
Then the candidate ECBF becomes
\begin{flalign}\label{eq:h}
    h(x) = d - (a \beta^2 + b \beta r + c r^2),
\end{flalign}
where $a$, $b$, $c$, and $d$ parameterize the control barrier function and these are all positive in our case.
One may show that ${L_gh(x)\equiv 0}$ but ${L_gL_fh(x)\neq0}$, ${\forall x\in \mathbb{R}^n}$, and thus
$h(x)$ is of relative degree 2.
That is differentiation is needed for the control inputs, i.e., the steering rate $\dot\delta$ and torque rate $\dot\tau$  to appear, and motivates the use of an ECBF of Sec.~\ref{sec:background}.
For the function \eqref{eq:h} we have
\begin{flalign}\label{eq:lfh}
    L_fh(x) = -(2a \beta \dot{\beta} + b \dot{\beta} r +  b \beta \dot{r} + 2c r \dot{r}),
\end{flalign}
where $\dot{\beta}$ and $\dot{r}$ are obtained from \eqref{eq:r}.
Differentiating yields
\begin{flalign}\label{eq:lfh2}
    & L^2_fh(x) + L_gL_fh(x)u
    \\
    &= -\big( 2a \dot{\beta}^2 + 2b \dot{\beta} \dot{r} + 2c \dot{r}^2 
    + \Ddot{\beta}(2a\beta + br) 
    + \Ddot{r}(b\beta + 2cr) \big), \nonumber
\end{flalign}
where the control inputs, $\dot\delta$ and $\dot\tau$, appear in $\Ddot{r}$ and $\Ddot{\beta}$.

\subsection{Quadratic Programming CBF Design} \label{sec:QP}

A control barrier function quadratic program (CBF-QP) can then be formulated to  prevent the driver from moving the vehicle to unrecoverable states.
We assume that the driver commands the desired steering angle $\delta_{\rm d}$ and the desired torque $\tau_{\rm d}$.
Then, applying \eqref{eq:esafety_cond_control} as the constraint one may utilize the ECBF framework established above to filter the driver input and ensure safety.

We construct the quadratic program
\begin{flalign}\label{eq:costCBFQP}
&\mathbf{u}_{\rm QP}(x) = \argmin_{\mathbf{u} \in \mathbb{R}^{m+1} } \frac{1}{2}\mathbf{u}^\top \mathbf{H} \mathbf{u} + \mathbf{F}^\top\mathbf{u}, 
\\
&{\rm s.t.} ~ L^2_fh(x) + L_gL_fh(x)u + p_0 h(x) + p_1 L_fh(x) +\epsilon \geq 0, \nonumber
\end{flalign}
where ${\mathbf{u} = [\,u^\top\ \epsilon\,]^\top  =[\, \dot\delta\  \dot\tau\  \epsilon\,]^\top}$ and $\epsilon$ is a slack variable used to ensure numeric feasibility and counteract uncertainties in the experiments; see \cite{CAVCBF, breeden2023robust, alan2025} for robustification of CBFs.
Here the characteristic equation of ${F-GP^\top}$ takes the form 
\begin{equation}
(\lambda+\alpha_0)(\lambda+\alpha_1) = \lambda^2 + p_1 \lambda + p_0 = 0,
\end{equation}
cf.~\eqref{eq:char}, yielding 
${p_0 =\alpha_0\alpha_1, ~ p_1 = \alpha_0 + \alpha_1}$.
To formulate the shared control approach, the matrices in \eqref{eq:costCBFQP} are given as
\begin{equation}
    \mathbf{H} = {\rm diag}([w_{\delta}, w_{\tau}, w_{\epsilon}]), 
    \quad 
    \mathbf{F} = -[\, w_{\delta} \dot\delta_{\rm d} \ w_{\tau} \dot\tau_{\rm d} \ 0\, ]^\top, 
\end{equation}
which, apart from the weights incorporate the steering rate $\dot\delta_{\rm d}$ and driving torque rate $\dot\tau_{\rm d}$ requested by the driver. 

We remark that the experimental vehicle's hardware interfaces through the steering angle $\delta$ and torque $\tau$ for measured signals and expected commands.
To address this, finite differences were used on the vehicle to obtain ${[\, \delta_{\rm QP} \ \tau_{\rm QP}\,]}$ from ${[\, \dot{\delta}_{\rm QP} \ \dot{\tau}_{\rm QP}\,]}$. Additionally, input constraints could be added to the QP, but in practice we found this was not necessary due to the fast actuation of the vehicle.

\begin{figure}[t]
    \vspace{0.6em}
    \includegraphics[width=0.8\columnwidth]{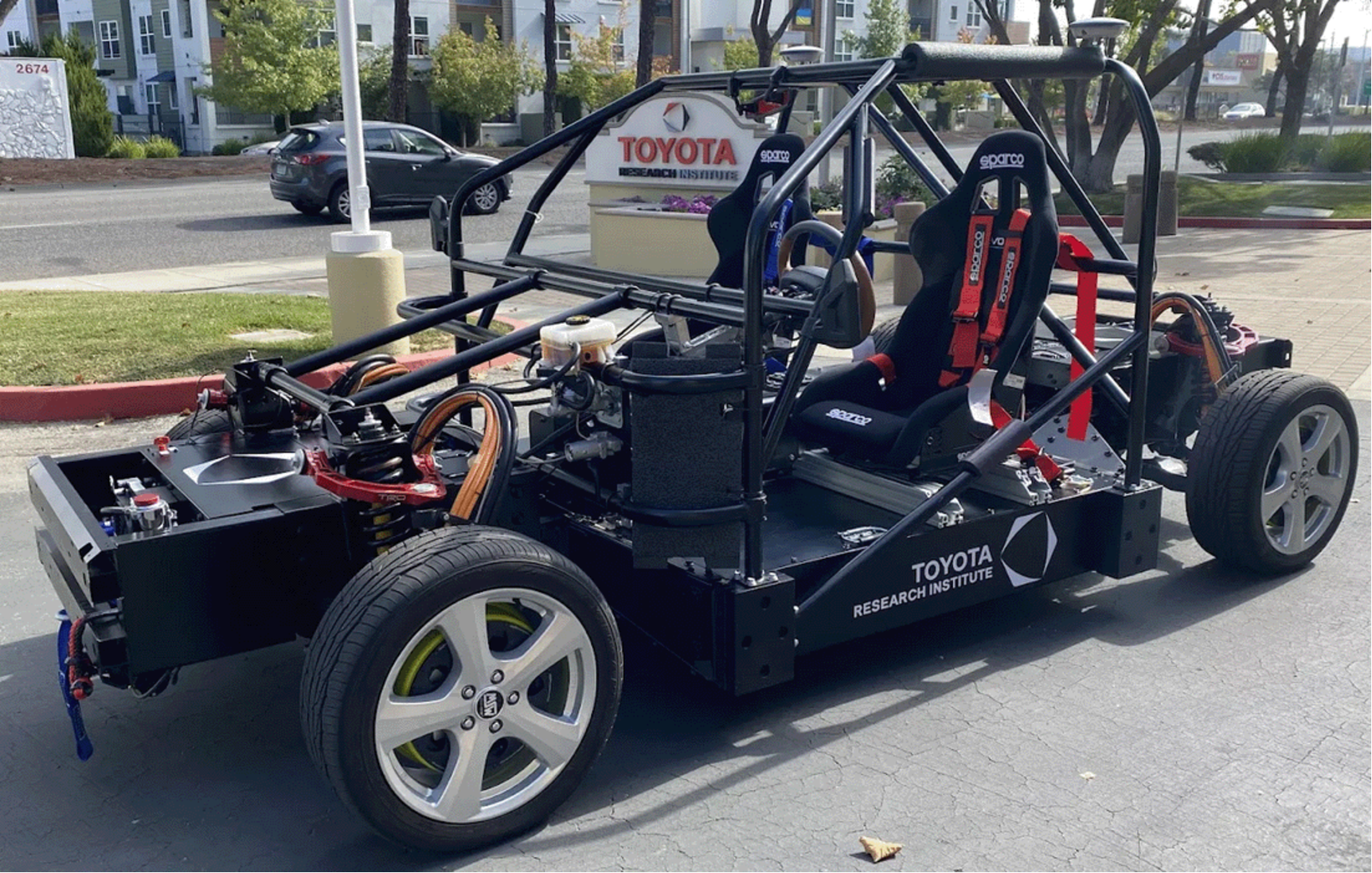}
    \centering
    \caption{Experimental platform.}
    \label{fig:GRIP}
        \vspace{-1.5em}
\end{figure}

\section{Experimental Validation} \label{sec:exp}

Here we present the experimental results that were obtained with a real vehicle on a test track. 
The vehicle was operated by a human driver whose inputs were filtered by the CBF based controller.
After describing the experimental platform we present simulation results which is followed by the experimental results.

\subsection{Experimental Platform}\label{sec:platform}

\begin{figure*}[t]
\vspace{0.2em}
    \centering
     \begin{subfigure}[t]{0.3\textwidth}
         \centering
         \includegraphics[scale = 0.28,trim={4cm 0 4cm 0}]{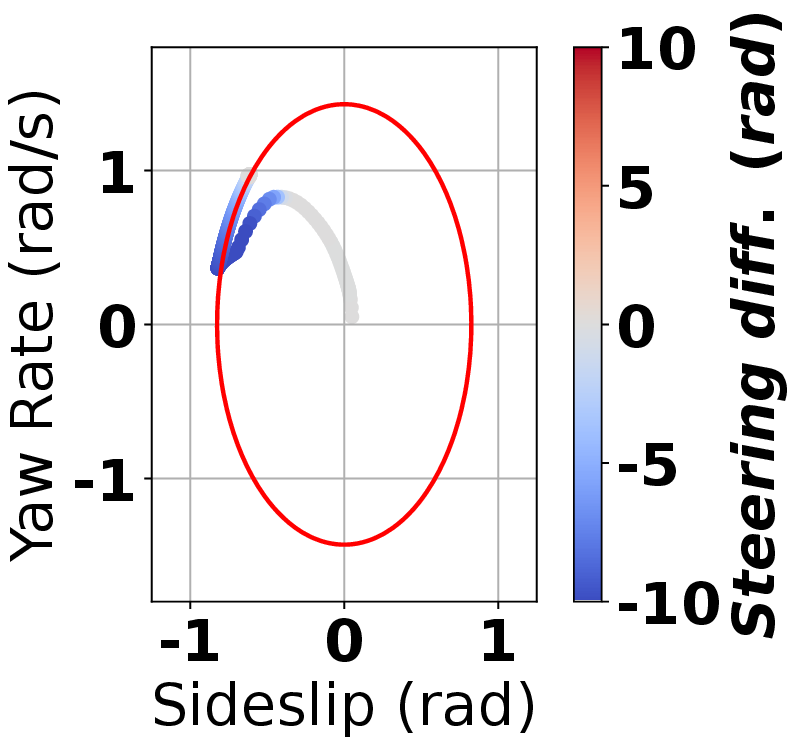}
         \caption{Drift initiation steering difference.}
         \label{fig: sim_init_delta}
     \end{subfigure}
     \begin{subfigure}[t]{0.3\textwidth}
         \centering
         \includegraphics[scale = 0.28,trim={4cm 0 4cm 0}]{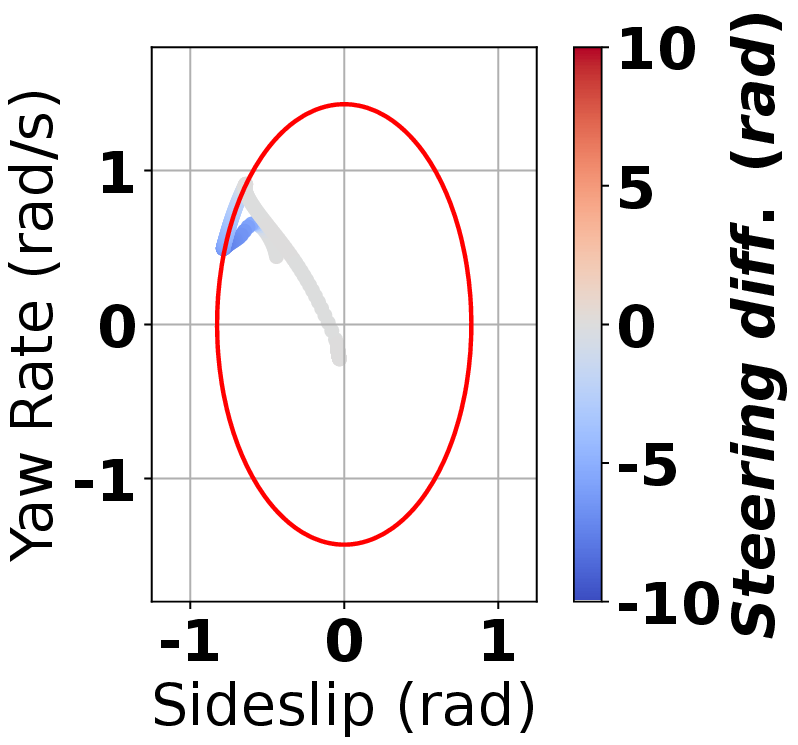}
         \caption{Drift equilibrium steering difference.}
         \label{fig: sim_eq_delta}
     \end{subfigure}
     \begin{subfigure}[t]{0.3\textwidth}
         \centering
         \includegraphics[scale = 0.28,trim={4cm 0 4cm 0}]{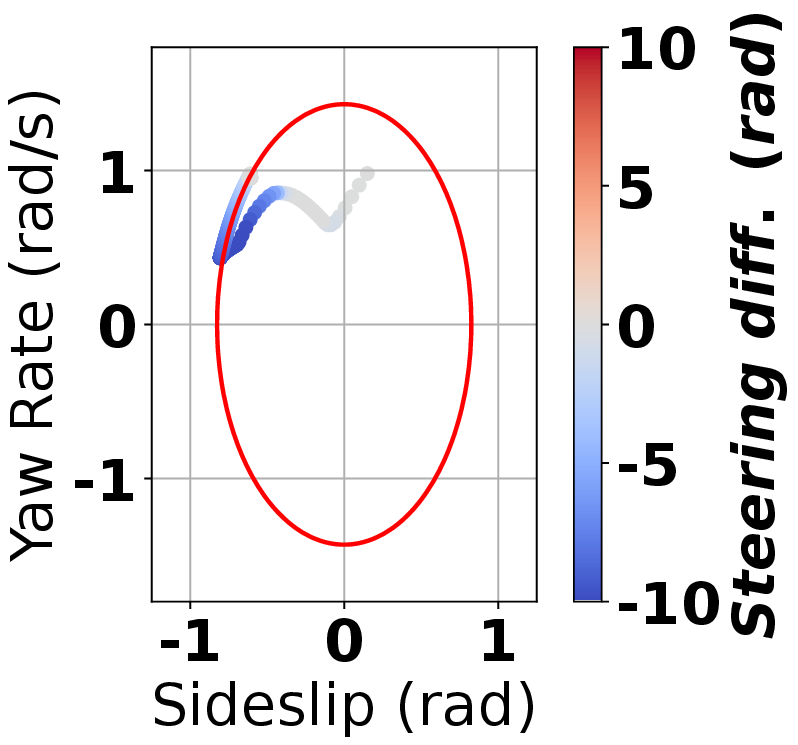}
         \caption{Drift transition steering difference.}
         \label{fig: sim_trans_delta}
     \end{subfigure}
     \begin{subfigure}[t]{0.3\textwidth}
         \centering
         \includegraphics[scale = 0.28, trim={4cm 0 4cm 0}]{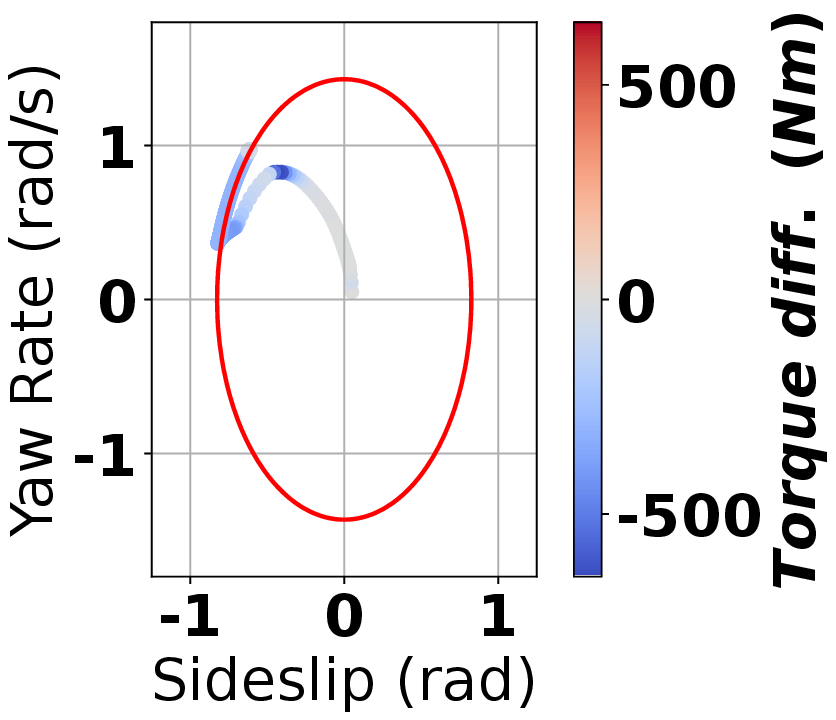}
         \caption{Drift initiation torque difference.}
         \label{fig: sim_init_tau}
     \end{subfigure}
     \begin{subfigure}[t]{0.3\textwidth}
         \centering
         \includegraphics[scale = 0.28, trim={4cm 0 4cm 0}]{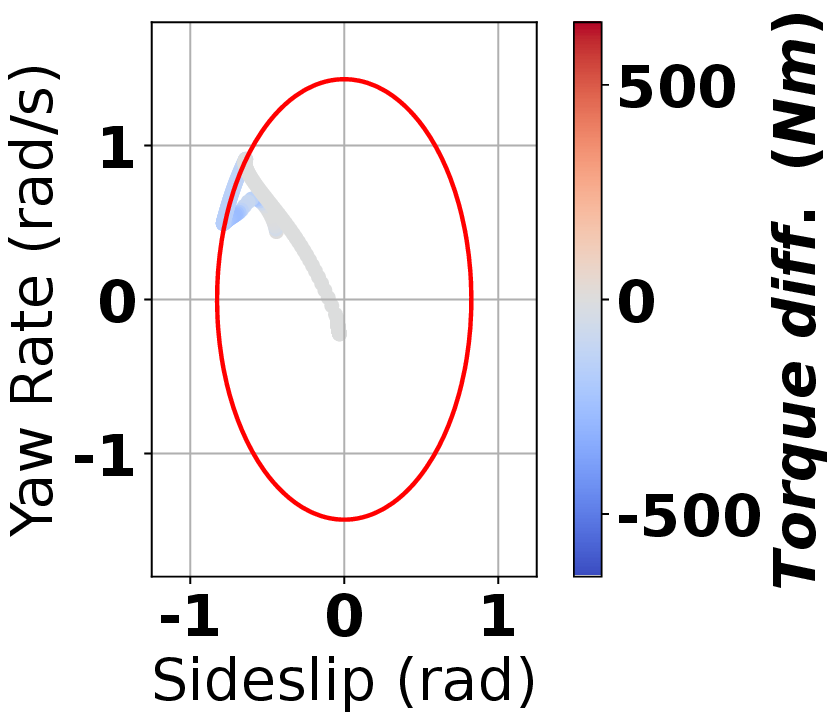}
         \caption{Drift equilibrium torque difference.}
         \label{fig: sim_eq_tau}
     \end{subfigure}
          \begin{subfigure}[t]{0.3\textwidth}
         \centering
         \includegraphics[scale = 0.28, trim={4cm 0 4cm 0}]{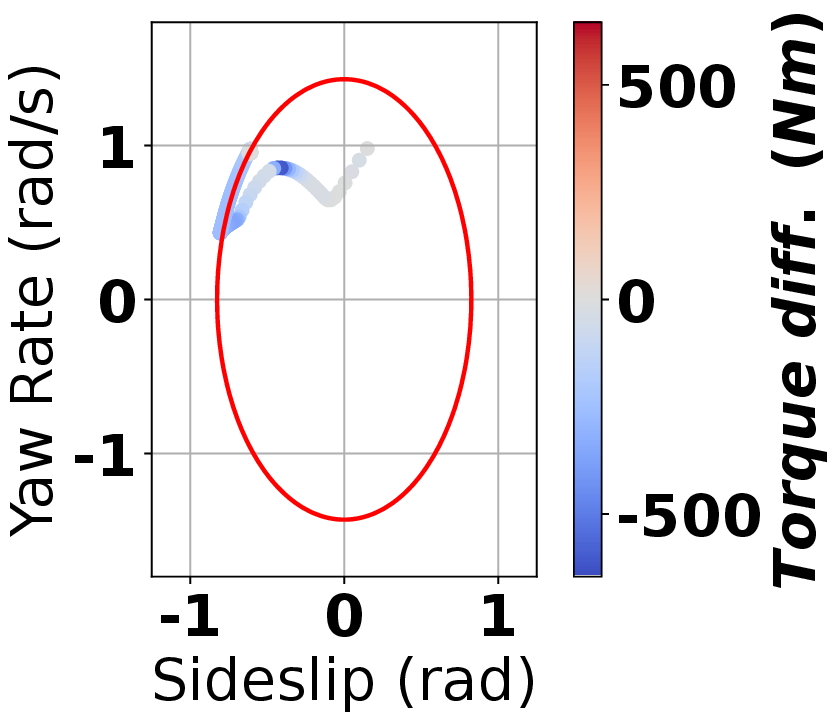}
         \caption{Drift transition torque difference.}
         \label{fig: sim_trans_tau}
     \end{subfigure}
    \caption{Phase plane plots for simulation validation. 
    The top row shows the difference between the steering angle commanded by the driver and realized by the vehicle after the safety filter is applied. 
    The bottom row shows this difference for the driving torque.}
    \label{fig:sim_init}
    \vspace{-1em}

\end{figure*}

Experiments are performed on a large skidpad.  The experimental vehicle (GRIP) is a custom made drive by wire platform shown in Fig.~\ref{fig:GRIP}. 
The vehicle has 4 independent in hub Elaphe IWM M700 VD4 motors, and two axle steering with a range of up to ${0.71\,\mathrm{rad}}$ at the front and rear axle. 
A dual antenna Oxford Technical Systems (OxTS) RT3003 RTK-GPS/IMU system provides state information at 200 Hz, and the CBF-QP is implemented on a RAVE ATC8110-F ruggedized computer with an Intel Xeon E-2278GE processor. 
All experiments were performed on a closed course.

The vehicle uses dynamics emulation \cite{russell} to exploit the extra degrees of freedom in the steering and in hub motors to replicate the dynamics of a vehicle driving on a low friction (${\mu=0.3}$) surface. 
Briefly, the inputs are mapped through a nonlinear single-track model with a coupled slip Fiala tire model parameterized by a friction coefficient of ${\mu = 0.3}$. 
This yields the desired yaw rate, sideslip angle, and speed of the emulated model. 
A dynamics inversion mapper then determines the combination of front and rear steering, independent motor torques, and independent brake torques, such that the physical vehicle's dynamics matches the emulated dynamics.


\subsection{Simulation Results}

Simulations are performed to demonstrate the ability of the CBF-QP of Sec.~\ref{sec:QP} to mitigate entering an unrecoverable part of the state space. 
Three shared control cases are considered: 1) the driver initiates a drift, 2) the driver attempts a counterclockwise stable drift, and 3) the driver attempts a transition in a figure 8 maneuver. 
These three cases assess various aspects of vehicle control. Specifically, 1) assesses transitions from grip driving to sliding, 2) assesses stabilizing a slide, and 3) assesses dynamic transitions. 
In all simulation cases, the driver applies excessive torque and reduced countersteer, motivating the use of the proposed safety filter as these inputs would lead to a loss of control. 
For brevity, we demonstrate the results for the upper half of the MPREl (counterclockwise drifting), but the simulations are also performed for the bottom half (clockwise drifting) and, as expected, yield a mirrored steering response. 
In all discussion, results are presented for the rear axle torque and steering angle at the handwheel. 
A linear map with a steering ratio of ${1/15}$ can be used to convert to roadwheel angle.

\subsubsection{Drift Initiation}

To test the drift initiation, the vehicle starts near the origin, while the driver applies a steering input of ${1.5\,\mathrm{rad}}$ (measured at the handwheel) and torque input ${856\,\mathrm{Nm}}$. 
Figs.~\ref{fig: sim_init_delta} and \ref{fig: sim_init_tau} show the $\beta-r$ phase plane with the difference between the CBF-QP commands and driver request for steering angle and torque, respectively. 
A grey color means the CBF is accepting the driver input, whereas a blue color means the CBF is reducing the driver input. 

The driver's commands cause the vehicle to evolve in a counterclockwise direction towards the boundary of the safe set (red ellipse), while the CBF-QP prevents the vehicle from losing control. 
Notably, the CBF-QP reduces the torque by a maximum of ${678\,\mathrm{Nm}}$ (at ${\beta=-0.41\,\mathrm{rad}}$) while the steering is augmented by ${-11.4\,\mathrm{rad}}$ at the handwheel. 
These actions prevent the vehicle from losing control as reducing the torque allows for a larger allocation of rear tire force to go to the stabilizing lateral force and increasing countersteer increases the lateral force on the front axle which balances the yaw moment of the vehicle, preventing a loss of control. 
The CBF begins to return to the driver commands as the vehicle evolves along the safe set boundary and stabilizes at a point around ${[\beta, ~ r] = [-0.61\,\mathrm{rad}, ~ 0.96\,\mathrm{rad/s}]}$. 
The mean solve time for simulation is ${38.7\, \mathrm{\mu s}}$, demonstrating real time applicability.

\subsubsection{Drift Equilibrium}

Next, the vehicle starts at a drift equilibrium ${[\beta, ~ r] = [-0.44\,\mathrm{rad}, ~  0.44\,\mathrm{rad/s}]}$, while the driver applies a steering input of ${-1.5\,\mathrm{rad}}$ and torque input ${700\,\mathrm{Nm}}$. 
The open-loop inputs to maintain a drift at this equilibrium are a steering input of ${-4.5\,\mathrm{rad}}$ and torque input ${364\,\mathrm{Nm}}$. 
Figs.~\ref{fig: sim_eq_delta} and \ref{fig: sim_eq_tau} show the difference between the CBF-QP commands and driver request for steering angle and torque, respectively. 

The vehicle evolves in a counterclockwise direction and the driver's commands cause the vehicle to evolve towards the boundary of the safe set (red ellipse). 
The CBF-QP prevents a loss of control by reducing the torque by ${272.6\,\mathrm{Nm}}$ (at ${\beta=-0.73\,\mathrm{rad}}$) while the steering is augmented by ${-10.3\,\mathrm{rad}}$ at the handwheel. 
Increasing countersteer and reducing torque in turn increases the lateral force on the the front and rear axles which balances the yaw moment of the vehicle, preventing a loss of control.  
The CBF-QP then begins to match the driver inputs as the trajectory progresses along the barrier boundary before returning toward the origin. 
The mean solve time for simulation is ${40\, \mathrm{\mu s}}$, demonstrating real time applicability.

\subsubsection{Drift Transition}

Next, the vehicle starts at a point corresponding to a figure 8 drift transition from a clockwise drift to a counterclockwise drift at ${[\beta, r] = [-0.98\,\mathrm{rad}, ~ 0.15\,\mathrm{rad/s}]}$, while the driver applies a steering input of ${1.5\,\mathrm{rad}}$ and torque input ${800\,\mathrm{Nm}}$. 
Figs.~\ref{fig: sim_trans_delta} and \ref{fig: sim_trans_tau} show the difference between the CBF-QP commands and driver request for steering and torque, respectively. 

The vehicle evolves in a counterclockwise direction and the driver's commands cause the vehicle to evolve towards the boundary of the safety set (red ellipse). 
The CBF-QP prevents the vehicle from losing control by reducing the torque by ${613\,\mathrm{Nm}}$ (at ${\beta=-0.38\,\mathrm{rad}}$) while the steering is augmented by ${-11\,\mathrm{rad}}$ at the handwheel.  
Increasing countersteer and reducing torque in turn increases the lateral force on the the front and rear axles which balances the yaw moment of the vehicle, preventing a loss of control. 
As the vehicle evolves along the boundary, the CBF-QP begins to return to matching the driver commands.
The mean solve time for simulation is ${39.8\, \mathrm{\mu s}}$, demonstrating real time applicability.

\subsection{Experimental Results}
\begin{figure*}[t]
\vspace{0.5em}
    \centering
     \begin{subfigure}[t]{0.45\textwidth}
         \centering
         \includegraphics[scale = 0.4,trim={4cm 0 4cm 0}]{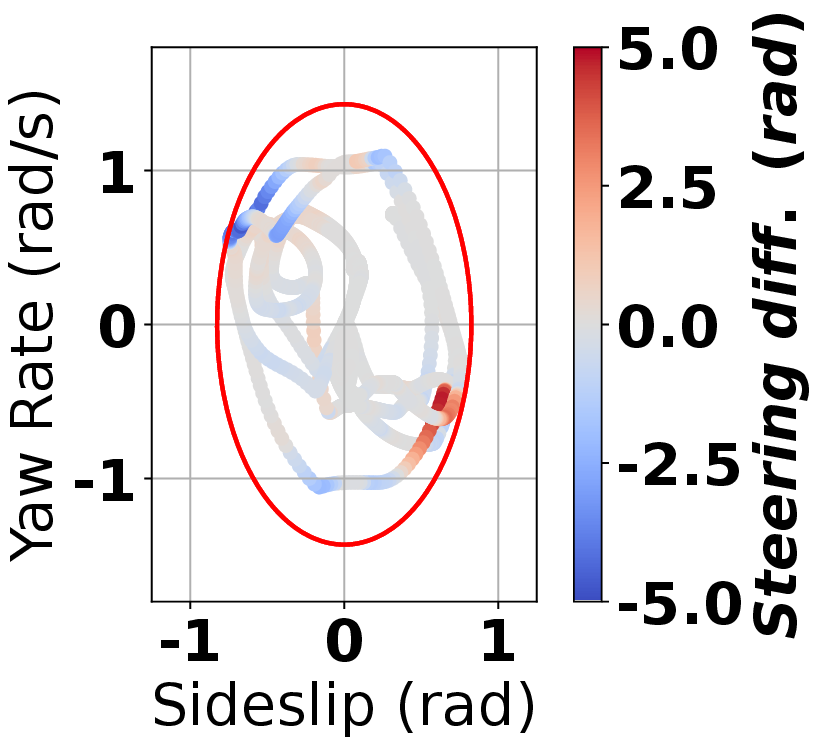}
         \caption{Experiment steering difference.}
         \label{fig:phase_portrait_delta}
     \end{subfigure}
     \begin{subfigure}[t]{0.45\textwidth}
         \centering
         \includegraphics[scale = 0.4, trim={3cm 0 5cm 0}]{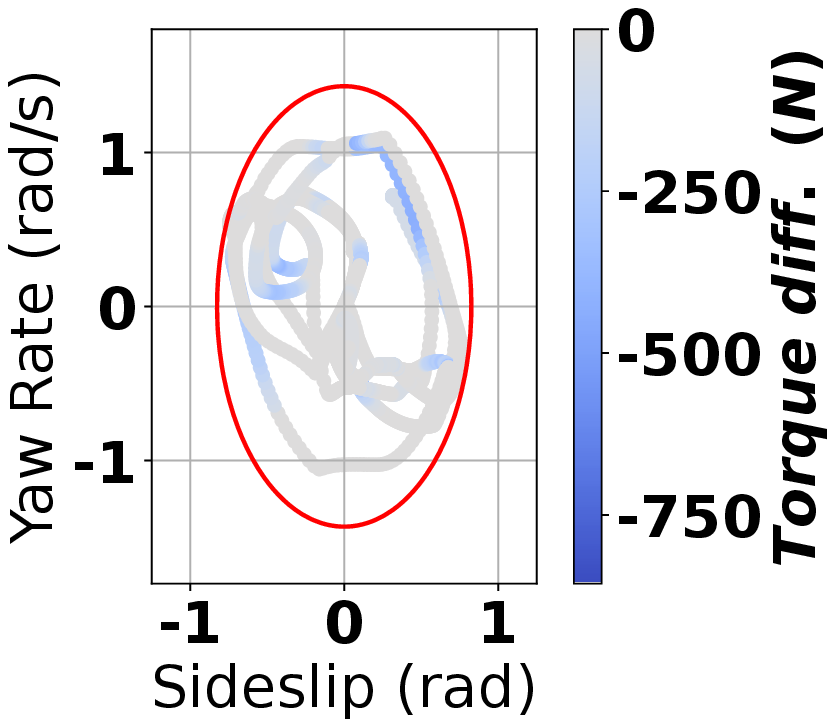}
         \caption{Experiment torque difference.}
         \label{fig:phase_portrait_throttle}
     \end{subfigure}
    \caption{Phase plane plots for experimental validation. 
    Panel (a) shows the difference between the steering angle commanded by the driver and realized by the vehicle after the safety filter is applied. 
    Panel (b) shows this difference for the driving torque.}
    \label{fig:exp_phase}
\end{figure*}

\begin{figure}[t]
    \includegraphics[scale = 0.22,trim={0cm 0 0cm 3cm}]{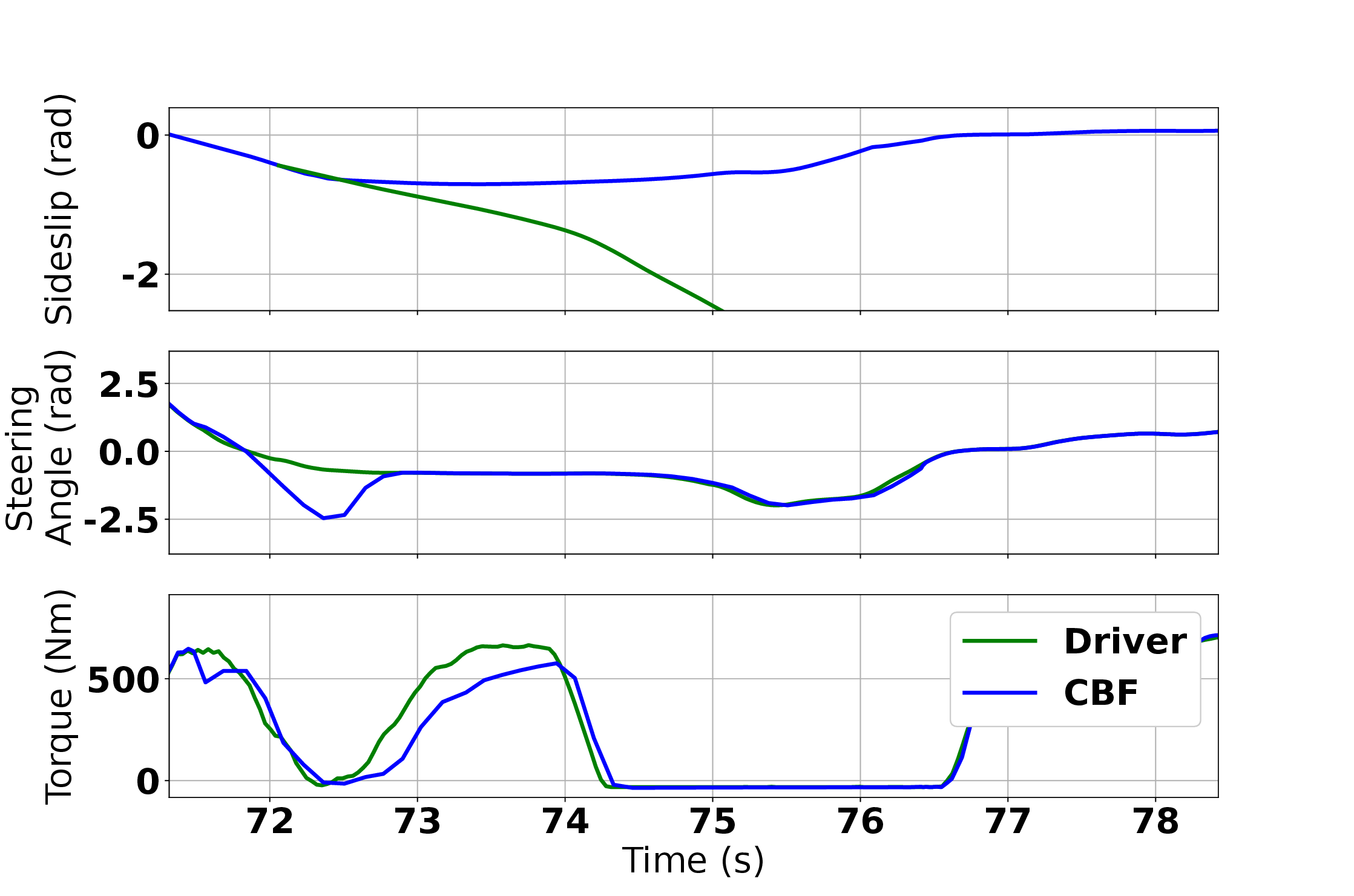}
    \centering
    \caption{Experimental trace for sideslip angle (top), steering angle (middle), and driving torque (bottom). 
    The CBF-QP output and resulting vehicle state are shown in blue while the requested driver input and simulated vehicle response are shown in green.}
    \label{fig:Experiment}
    \vspace{-1.5em}
\end{figure}

Next, the experiments performed on the GRIP platform are described for minimally invasive shared control.
As depicted in Fig.~\ref{fig:exp_overlay}, the driver is performing circular clockwise, counterclockwise, and transition drifts. 
The results are shown in Figs.~\ref{fig:phase_portrait_delta} and \ref{fig:phase_portrait_throttle} where the same notation is used as in Fig.~\ref{fig:sim_init}.

The trajectory evolves in counterclockwise loops, and the phase portrait can be divided into four parts. 
In quadrant~I (${\beta>0, ~ r>0}$) the vehicle is transitioning from a clockwise to counterclockwise drift (e.g., figure 8 transition). 
In quadrant~II (${\beta<0, ~ r>0}$) the vehicle is performing a counterclockwise drift. 
In quadrant~III (${\beta<0, ~ r<0}$) the vehicle is transitioning from a counterclockwise to clockwise drift. 
Finally, in quadrant~IV (${\beta>0, ~ r<0}$) the vehicle is performing a clockwise drift. 

Different aspects of the CBF-QP behavior are exhibited throughout the experiment.  
The small counterclockwise loops contained fully in quadrant~II represent a stable counterclockwise drift.   
As the driver is drifting in a stable manner, minimal intervention occurs in the steering (Fig. \ref{fig:phase_portrait_delta}) and torque (Fig. \ref{fig:phase_portrait_throttle}) when far from the boundary. 
However, as the vehicle approaches the boundary the CBF-QP augments countersteer by up to ${-5.5\,\mathrm{rad}}$ and reduces torque by ${226\,\mathrm{Nm}}$ to prevent a breach of safety. 

The larger loops that transition from quadrant~II (counterclockwise drift) through quadrant~III and to quadrant~IV (clockwise drift) occur during transitions from counterclockwise to clockwise loops.  
A successful transition often requires the torque to be released and the steering to transition from a negative countersteer to a positive countersteer, which is needed to stabilize the clockwise drift. 
These behaviors are observed where a reduction in torque of ${323\,\mathrm{Nm}}$ first occurs in Fig.~\ref{fig:phase_portrait_throttle} around ${[\beta = -0.62\,\mathrm{rad}, ~ r= -0.2\,\mathrm{rad/s}]}$, as the driver was applying maximum torque at this point. 
The torque then gradually returns to the driver commands and the CBF-QP helps transition the driver from negative countersteer to positive countersteer. 
Particularly, around ${[\beta = 0.65\,\mathrm{rad}, ~ r = -0.5\,\mathrm{rad/s}]}$ the CBF-QP intervenes through augmenting steering at the handwheel by ${5.1\,\mathrm{rad}}$, guiding the driver in this steering transition from negative countersteer to positive. 
Next, as the vehicle transitions from clockwise to counterclockwise drifts the phase portrait transitions from quadrant~IV (clockwise drift) through quadrant~I to quadrant~II (counterclockwise drift). 
A similar behavior is observed where a large torque reduction of ${407\,\mathrm{Nm}}$ occurs at ${[\beta = 0.41\,\mathrm{rad}, ~ r = 0.71\,\mathrm{rad/s}]}$, followed by a steering augmentation of ${-5.5\,\mathrm{rad}}$ of countersteer at ${[\beta = -0.65\,\mathrm{rad}, ~ r = 0.6\,\mathrm{rad/s}]}$. 
These actions save the vehicle from breaching the barrier in the upper left corner which would lead to a loss of safety.

\begin{figure}[ht]
    \includegraphics[scale = 0.25,trim={0cm 0 0cm 2cm}]{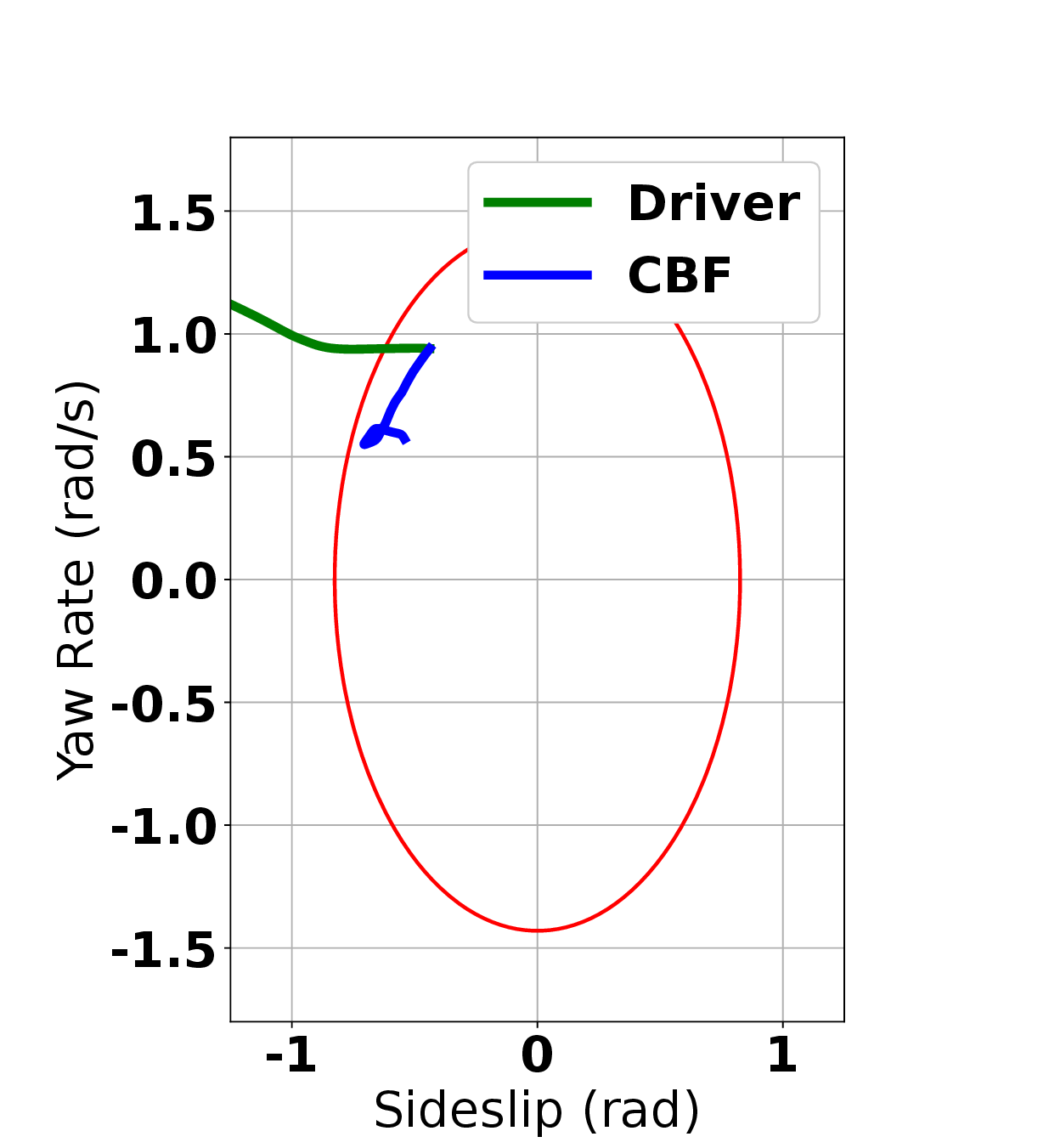}
    \centering
    \caption{Trajectory from CBF-QP experiment (blue) and forward simulation of model using the driver's inputs (green).}
    \label{fig:Experiment_pp}
    \vspace{-1.5em}
\end{figure}

To highlight the impact of these interventions,  Fig.~\ref{fig:Experiment} shows the CBF-QP commands in blue and the measured driver inputs in green (bottom two plots). 
A forward simulation of sideslip assuming the driver's commands were applied directly is shown in green (top) starting from ${[\beta, ~ r] = [-0.3\,\mathrm{rad}, ~ 1.04\,\mathrm{rad/s}]}$. 
The CBF-QP follows the driver's intentions in a safe manner: as the driver adds or removes torque, so does the CBF-QP, and similarly for the steering input. 
This consistency among the driver and CBF-QP commands ensures the drivers intentions are matched, in this case allowing for a transition from positive to negative sideslip (i.e., a figure 8 transition) just as the driver intends. 
However, not only does the CBF-QP follow the driver intentions, but it does so in a safe manner by significantly reducing the driver torque by ${210\,\mathrm{Nm}}$ (bottom) and augmenting countersteering at the handwheel by ${-1.85\,\mathrm{rad}}$ (middle) from ${t=72}$ to ${74\, \mathrm{s}}$. These actions maintain stability of the vehicle as shown by the measured sideslip not exceeding ${-0.71\,\mathrm{rad}}$. In contrast a post-hoc simulation which applies the driver's inputs to the single track model of Sec.~\ref{sec:approach} leads to a large sideslip indicative of a spin out. 
Once the vehicle is in a safe state (${74.4\, \mathrm{s}}$), the CBF-QP begins to match the driver inputs. 
This is also highlighted in Fig.~\ref{fig:Experiment_pp} where the CBF-QP (blue) evolves in a safe manner whereas the unfiltered driver inputs (green) lead to a breach of the safety. 
These results demonstrate the utility of proposed approach to maintain vehicle stability in real world experiments.

\section{Conclusion} \label{sec:conc}

This paper presented a minimally invasive shared control approach to maintaining vehicle safety in a computationally efficient manner by leveraging exponential control barrier functions. 
The maximal phase recoverable ellipse was utilized as a safe set in the sideslip angle--yaw rate phase plane and an ECBF was placed on the MPREl. 
Simulations demonstrated the CBF-QP at various drifting maneuvers, and experiments performed on a full scale vehicle validated the approach in a shared control setting for figure 8 drifting maneuvers. 
These experiments demonstrate the CBF-QP matches driver inputs if safe, and, if needed, augments steering and torque to balance the yaw moment of the vehicle and prevent a breach of the barrier. 
The designed safety filter prevents a vehicle from entering unrecoverable states with a mean computation time less than ${50\ \mathrm{\mu s}}$. 
These results represent an important step in transitioning research on autonomous vehicles operating at the limits of their capabilities to broader, potentially production-oriented, designs.

Future work could improve the robustness of the safety filter by explicitly accounting for or adapting to model uncertainty in the CBF formulation. Future work could also explore the impact of actuation constraints, although we found this work achieved good performance despite this thanks to the fast actuators on the vehicle.

\bibliographystyle{IEEEtran}
\bibliography{main}

\begin{thebibliography}{10}
\providecommand{\url}[1]{#1}
\csname url@samestyle\endcsname
\providecommand{\newblock}{\relax}
\providecommand{\bibinfo}[2]{#2}
\providecommand{\BIBentrySTDinterwordspacing}{\spaceskip=0pt\relax}
\providecommand{\BIBentryALTinterwordstretchfactor}{4}
\providecommand{\BIBentryALTinterwordspacing}{\spaceskip=\fontdimen2\font plus
\BIBentryALTinterwordstretchfactor\fontdimen3\font minus \fontdimen4\font\relax}
\providecommand{\BIBforeignlanguage}[2]{{%
\expandafter\ifx\csname l@#1\endcsname\relax
\typeout{** WARNING: IEEEtran.bst: No hyphenation pattern has been}%
\typeout{** loaded for the language `#1'. Using the pattern for}%
\typeout{** the default language instead.}%
\else
\language=\csname l@#1\endcsname
\fi
#2}}
\providecommand{\BIBdecl}{\relax}
\BIBdecl

\bibitem{dallas2023hierarchical}
M.~Thompson, J.~Dallas, J.~Y.~M. Goh, and A.~Balachandran, ``Adaptive nonlinear model predictive control: Maximizing tire force and obstacle avoidance in autonomous vehicles,'' \emph{IEEE Transactions on Field Robotics}, vol.~1, pp. 318--331, 2024.

\bibitem{LaurenseThesis}
V.~Laurense, ``Integrated motion planning and control for automated vehicles up to the limits of handling,'' Ph.D. dissertation, Stanford University, 2019.

\bibitem{Goh2023}
J.~Y.~M. Goh, M.~Thompson, J.~Dallas, and A.~Balachandran, ``Beyond the stable handling limits: nonlinear model predictive control for highly transient autonomous drifting,'' \emph{Vehicle System Dynamics}, vol.~62, no.~10, pp. 1--24, 2024.

\bibitem{talbot2023shared}
J.~Talbot, M.~Brown, and J.~C. Gerdes, ``Shared control up to the limits of vehicle handling,'' \emph{IEEE Transactions on Intelligent Vehicles}, 2023.

\bibitem{SMCDrift}
I.~Karino, J.~Dallas, and J.~Y.~M. Goh, ``Shared control for giving ordinary drivers expert level drifting skills,'' in \emph{IEEE International Conference on Systems, Man, and Cybernetics (SMC)}, 2023, pp. 1461--1467.

\bibitem{ames2019control}
A.~D. Ames, S.~Coogan, M.~Egerstedt, G.~Notomista, K.~Sreenath, and P.~Tabuada, ``Control barrier functions: Theory and applications,'' in \emph{European Control Conference (ECC)}, 2019, pp. 3420--3431.

\bibitem{XU201554}
X.~Xu, P.~Tabuada, J.~W. Grizzle, and A.~D. Ames, ``Robustness of control barrier functions for safety critical control,'' \emph{IFAC-PapersOnLine}, vol.~48, no.~27, pp. 54--61, 2015, {C}onference on Analysis and Design of Hybrid Systems (ADHS).

\bibitem{Ames2017}
A.~D. Ames, X.~Xu, J.~W. Grizzle, and P.~Tabuada, ``Control barrier function based quadratic programs for safety critical systems,'' \emph{IEEE Transactions on Automatic Control}, vol.~62, pp. 3861--3876, 2017.

\bibitem{ECBF}
Q.~Nguyen and K.~Sreenath, ``Exponential control barrier functions for enforcing high relative-degree safety-critical constraints,'' in \emph{American Control Conference (ACC)}, 2016, pp. 322--328.

\bibitem{cbfcomp}
J.~Breeden and D.~Panagou, ``Compositions of multiple control barrier functions under input constraints,'' in \emph{American Control Conference (ACC)}, 2023, pp. 3688--3695.

\bibitem{CBFMPC}
J.~Zeng, B.~Zhang, and K.~Sreenath, ``Safety-critical model predictive control with discrete-time control barrier function,'' in \emph{American Control Conference (ACC)}, 2021, pp. 3882--3889.

\bibitem{AVEC2024}
C.~Jiang, H.~Gan, I.~V{\"o}r{\"o}s, D.~Tak{\'a}cs, and G.~Orosz, ``Safety filter for lane-keeping control,'' in \emph{16th International Symposium on Advanced Vehicle Control}.\hskip 1em plus 0.5em minus 0.4em\relax Springer, 2024, pp. 371--377.

\bibitem{CAVCBF}
A.~Alan, A.~J. Taylor, C.~R. He, A.~D. Ames, and G.~Orosz, ``Control barrier functions and input-to-state safety with application to automated vehicles,'' \emph{IEEE Transactions on Control Systems Technology}, vol.~31, no.~6, pp. 2744--2759, 2023.

\bibitem{goh2019automated}
J.~Y.~M. Goh, \emph{Automated Vehicle Control Beyond the Stability Limits}.\hskip 1em plus 0.5em minus 0.4em\relax Stanford University, 2019.

\bibitem{cohen2024safety}
M.~H. Cohen, T.~G. Molnar, and A.~D. Ames, ``Safety-critical control for autonomous systems: Control barrier functions via reduced-order models,'' \emph{Annual Reviews in Control}, vol.~57, p. 100947, 2024.

\bibitem{XIAO2021109592}
W.~Xiao, C.~G. Cassandras, and C.~A. Belta, ``Bridging the gap between optimal trajectory planning and safety-critical control with applications to autonomous vehicles,'' \emph{Automatica}, vol. 129, p. 109592, 2021.

\bibitem{molnar2023safety}
T.~G. Molnar, G.~Orosz, and A.~D. Ames, ``On the safety of connected cruise control: analysis and synthesis with control barrier functions,'' in \emph{62nd IEEE Conference on Decision and Control (CDC)}.\hskip 1em plus 0.5em minus 0.4em\relax IEEE, 2023, pp. 1106--1111.

\bibitem{Xiao2019}
W.~Xiao and C.~Belta, ``Control barrier functions for systems with high relative degree,'' in \emph{58th IEEE Conference on Decision and Control (CDC)}, 2019, pp. 474--479.

\bibitem{oh2021handling}
S.~Oh, S.~S. Avedisov, and G.~Orosz, ``On the handling of automated vehicles: modeling, bifurcation analysis, and experiments,'' \emph{European Journal of Mechanics A}, vol.~90, p. 104372, 2021.

\bibitem{CombinedCBF}
T.~G. Molnar and A.~D. Ames, ``Composing control barrier functions for complex safety specifications,'' \emph{IEEE Control Systems Letters}, vol.~7, pp. 3615--3620, 2023.

\bibitem{DallasMPRP}
J.~Dallas, I.~Karino, M.~Thompson, B.~Araki, S.~Goldine, J.~Y.~M. Goh, and J.~Subosits, ``Safe stability envelopes and shared control for active vehicle safety,'' in \emph{IEEE Conference on Control Technology and Applications (CCTA)}, 2024, pp. 561--568.

\bibitem{breeden2023robust}
J.~Breeden and D.~Panagou, ``Robust control barrier functions under high relative degree and input constraints for satellite trajectories,'' \emph{Automatica}, vol. 155, p. 111109, 2023.

\bibitem{alan2025}
A.~Alan, T.~G. Moln\'ar, A.~D. Ames, and G.~Orosz, ``Generalizing robust control barrier functions from a controller design perspective,'' \emph{IEEE Open Journal of Control Systems}, vol.~4, pp. 54--69, 2025.

\bibitem{russell}
H.~E.~B. Russell and J.~C. Gerdes, ``Design of variable vehicle handling characteristics using four-wheel steer-by-wire,'' \emph{IEEE Transactions on Control Systems Technology}, vol.~24, no.~5, pp. 1529--1540, 2016.

\end{thebibliography}

\end{document}